\documentclass[prl, twocolumn,superscriptaddress,amsmath,amssymb,showpacs,floatfix,preprintnumbers]{revtex4-2}

\usepackage{amsmath}
\usepackage{graphicx}
\usepackage{dcolumn}
\usepackage{xcolor}
\usepackage{physics}
\usepackage{xr}
\usepackage{siunitx}
\usepackage{comment}
\usepackage{tikz}
\usepackage{tabularx}
\usepackage{makecell}
\usepackage{multirow}
\usepackage{float}
\usepackage{amssymb}
\DeclareGraphicsExtensions{.png .jpg .pdf}

\begin{document}

\title{Metallic electro-optic effects in topological chiral crystals}
\author{C. O. Ascencio}
\affiliation{School of Physics and Astronomy, University of Minnesota, Minneapolis, Minnesota 55455, USA}
\author{D. J. P. de Sousa}
\affiliation{Department of Electrical and Computer Engineering, University of Minnesota, Minneapolis, Minnesota 55455, USA}
\author{Tony Low}\email{tlow@umn.edu}
\affiliation{School of Physics and Astronomy, University of Minnesota, Minneapolis, Minnesota 55455, USA}

\affiliation{Department of Electrical and Computer Engineering, University of Minnesota, Minneapolis, Minnesota 55455, USA}

\date{ \today }

\begin{abstract}
Topological chiral crystals have emerged as a fertile material platform for investigating optical phenomena derived from the distinctive Fermi surface Berry curvature and orbital magnetic moment textures around multifold chiral band crossings pinned at the time-reversal invariant momenta. In this work, by means of tight-binding model and first principles based calculations, we investigate metallic electro-optic (EO) responses stemming from the Berry curvature and orbital magnetic moment of Bloch electrons across 37 materials belonging to space group 198 (SG198). Previously thought to vanish in SG198, our findings reveal a nonzero Berry curvature dipole attributed to the energetic misalignment between topologically charged point nodes of opposite chirality. Moreover, we find that the recently predicted magnetoelectric EO effects, which arise from the interplay between the Berry curvature and magnetic moment on the Fermi surface, are readily accessible in BeAu under experimentally feasible electric biases.
\end{abstract}

\maketitle

\section{I. Introduction}
 Berry curvature enabled electron responses in quantum materials have been the subject of intensive research for well over a decade~\cite{QHE,Haldane, PhysRevLett.97.026603, PhysRevLett.105.026805, PhysRevB.103.L201202,Farajollahpour2025, AHE, AHE2, SHE, QSHE, VHE2, VHE, THE, NLHE, SHG, BCD_lasing, Ivo_LEO, Te_BCD_gain, BCD_transistor, Duarte_BCD}. These research efforts have led to the discovery of a family of anomalous Hall effects~\cite{Haldane, AHE, AHE2, SHE, QSHE, VHE2, VHE, THE, NLHE, Tony_paper} as well as emerging optoelectronic phenomena~\cite{Kim2019,doi:10.1126/sciadv.adk3897, BCD_lasing, Ivo_LEO, Te_BCD_gain, BCD_transistor}. Previous work on the intraband electro-optic (EO) properties of time-reversal invariant, non-centrosymmetric metals were primarily concerned with  Berry curvature dipole (BCD) mediated responses, which emerge from the Fermi surface Berry curvature distribution~\cite{Ivo_LEO,BCD_transistor,Kim2019}. The BCD has been theoretically predicted to give rise to second harmonic optical responses, non-Hermitian EO effects, and is touted to originate nonreciprocal optical gain~\cite{SHG, BCD_lasing, Ivo_LEO, Te_BCD_gain, BCD_transistor, NLHE, Tony_paper}. Importantly, recent studies have identified a unique class of Berry curvature-driven metallic EO effects that are magnetoelectric in nature, originating from the coupling between the Fermi surface magnetic moment and Berry curvature of Bloch electrons in systems lacking either inversion or time-reversal symmetry~\cite{magnetoelectric_electro-optical}. However, these magnetoelectric EO effects remain largely~\cite{MEEO_TDBG} unexplored in real materials and identifying an ideal material platform to realize them could unlock new opportunities for optical control and tunable light–matter interactions.
 
Nonmagnetic chiral crystals constitute a class of noncentrosymmetric topological metals with rich Berry curvature and orbital magnetic moment phenomenology, providing an ideal platform for exploring newly predicted metallic EO effects in the terahertz regime~\cite{chiral_crystals, RhSi,Ivo_Te, chiral_optical, pnas_Felser, Felser2}. As a consequence of their structural chirality and time-reversal symmetry, point-like topologically charged nodes may appear in their band structures at the time-reversal invariant momenta (TRIM) when spin-orbit coupling (SOC) is considered~\cite{chiral_crystals}. Importantly, topological nodes of opposite chirality (Chern number) pinned at the TRIMs are not related by symmetry, so can occur with an energy offset~\cite{RhSi, chiral_crystals, Pshenay-Severin_2018}. Time-reversal symmetric crystals from space group 198 (SG198) have been shown to support topological highfold fermionic linear band crossings which carry the largest possible Chern number achievable by a zero dimensional node~\cite{Slager,Bradlyn2,Bradlyn2017, chiral_crystals, Pshenay-Severin_2018, RhSi}. The bands forming these highfold fermions have Fermi surfaces  with Berry curvature and orbital moment textures resembling a real-space magnetic monopole~\cite{chiral_crystals,pnas_Felser} of which the latter texture has been predicted to enable large orbital Hall and orbital magnetoelectric  responses~\cite{pnas_Felser}. Additionally, quantum states near topologically charged highfold nodes are strong sources Berry curvature and orbital moment~\cite{Felser2, pnas_Felser, SG198_SHE}, which can significantly contribute to metallic EO effects. Taken together, these properties underscore the potential of SG198 chiral crystals to host unconventional metallic EO phenomena.

In this work, we study metallic EO effects in topological chiral metals pertaining to SG198. By means of a tight-binding model and first principles based calculations, we evaluate the optical response tensors describing EO effects associated with Fermi surface Berry curvature and orbital magnetic moment textures. We investigate their energy dependence and accurately describe the observed features around the highfold topological nodes using symmetry arguments and available $\textbf{k} \cdot \textbf{p}$ models. Previously thought to vanish in SG198 crystals, we predict a nonzero BCD which exists as a consequence of a Fermi sea net Berry curvature divergence due to the energetic misalignment between topological nodes of opposite chirality pinned at the TRIMs. Based on first principles calculations of 37 experimentally observed nonmagnetic SG198 materials, we identify BeAu~\cite{BeAu_1,BeAu_2} as a promising material candidate  for experimentally investigating the metallic magnetoelectric EO effects considered in this work.

\newpage
The manuscript is organized as follows: In Sec. II we briefly summarize the formalism used to describe the current response to optical fields in the presence of an electric bias for noncentrosymmetric systems with time-reversal symmetry and define relevant optical response tensors. A description of our tight-binding model and an analysis of the calculated energy-dependent optical response tensors is given in Sec. III. A comparison of metallic EO effects derived from the BCD and the recently predicted magnetoelectric EO tensor~\cite{magnetoelectric_electro-optical} between realistic SG198 materials based on first principles calculations is the subject of Sec. IV.

\section{II. Formalism}

In this section, we briefly summarize the formalism employed in this work. We refer the interested reader to the Ref.~\cite{magnetoelectric_electro-optical} for more details.

\subsection{A. Constitutive relations in the presence of bias}

We begin by summarizing the basic constitutive relations in the absence of bias field, $\textbf{E}_0 = \textbf{0}$. We adopt the relaxation-time approximation within the semiclassical Boltzmann formalism to compute the nonequilibrium distribution function driven by monochromatic optical fields $\textbf{E}(\omega) = \textbf{E}_{\omega}e^{-i\omega t} + \textbf{E}^{*}_{\omega}e^{i\omega t}$ and $\textbf{B}(\omega) = \textbf{B}_{\omega}e^{-i\omega t} + \textbf{B}^{*}_{\omega}e^{i\omega t}$. The constitutive relation for time-reversal symmetric systems is
\begin{eqnarray}
J^{\alpha}(\omega) = \sigma^{\alpha\beta}_{E}(\omega)E_{\omega}^{\beta}  +  \sigma^{\alpha\beta}_{B}(\omega) B^{\beta}_{\omega}.\nonumber \\
    \label{Eq1}
\end{eqnarray}
The transport coefficients include the AC Drude conductivity
\begin{eqnarray}
\sigma_{E}^{\alpha\beta}(\omega) = \displaystyle \frac{e^2\tau}{1 - i\omega\tau}\sum_n \int \frac{d^3\textbf{k}}{(2\pi)^3}\left(-\frac{\partial f_{n\textbf{k}}^0}{\partial \epsilon_{n\textbf{k}}}\right)v_{n\textbf{k}}^{\alpha}v_{n\textbf{k}}^{\beta}, 
    \label{Eq2}
\end{eqnarray}
where $f_{n\textbf{k}}^0$ is the equilibrium Fermi-Dirac distribution, $\textbf{v}_{n\textbf{k}}$ is the Bloch electron velocity, and the dynamical magnetoelectric coupling coefficient~\cite{Ivo_gyro}
\begin{eqnarray}
\sigma_{B}^{\alpha\beta}(\omega) = \displaystyle e \frac{i\omega\tau}{i\omega\tau - 1}\sum_n \int \frac{d^3\textbf{k}}{(2\pi)^3}\left(-\frac{\partial f_{n\textbf{k}}^0}{\partial \epsilon_{n\textbf{k}}}\right)v_{n\textbf{k}}^{\alpha}m_{n\textbf{k}}^{\beta}, \nonumber \\
    \label{Eq3}
\end{eqnarray}
where $\textbf{m}_{n\textbf{k}}$ is the magnetic moment of Bloch electrons in momentum space, including orbital and spin contributions $\textbf{m}_{n\textbf{k}} = \textbf{m}^{\textrm{orbital}}_{n\textbf{k}} + \textbf{m}^{\textrm{spin}}_{n\textbf{k}}$. The orbital contribution reads~\cite{Ivo_gyro, PhysRevLett.99.197202, PhysRevB.94.245121, RevModPhys.82.1959}  
\begin{eqnarray}
    \textbf{m}^{\textrm{orb}}_{n\textbf{k}} = \displaystyle \frac{e}{2\hbar} \operatorname{Im}\langle \grad_{\textbf{k}} u_{n\textbf{k}} | \times (H_{\textbf{k}} - \epsilon_{n\textbf{k}})| \grad_{\textbf{k}} u_{n\textbf{k}}\rangle,
    \label{eq}
\end{eqnarray}
while the spin contribution is $\textbf{m}^{\textrm{spin}}_{n\textbf{k}} = -(eg_s\hbar/4m) \langle u_{n\textbf{k}}| \boldsymbol{\sigma} | u_{n\textbf{k}}\rangle$, where $g_s$ is the spin g-factor and $m$ the electron mass.

The presence of a static electric field, $\textbf{E}_0$, modifies the electromagnetic properties of the medium, giving rise to EO effects~\cite{Ivo_LEO, magnetoelectric_electro-optical, BCD_transistor}. In this work, we describe EO effects originating from the two possible corrections to the conductivities, Eqs.~(\ref{Eq2}) and (\ref{Eq3}), induced by $\textbf{E}_0$, namely the anomalous velocity and the correction to the distribution function. Explicitly,
\begin{eqnarray}
\textbf{v}_{n\textbf{k}} \rightarrow \textbf{v}_{n\textbf{k}} + \frac{e}{\hbar}(\textbf{E}_0\times\boldsymbol{\Omega}_{n\textbf{k}}), 
    \label{Eq4}
\end{eqnarray}
\begin{eqnarray}
f_{n\textbf{k}}^0 \rightarrow \displaystyle f_{n\textbf{k}}^0 + e\tau \frac{\partial f_{n\textbf{k}}^0}{\partial \epsilon_{n\textbf{k}}}(\textbf{v}_{n\textbf{k}}\cdot\textbf{E}_0),
    \label{Eq5}
\end{eqnarray}
respectively. The new contributions are necessarily Fermi surface terms, which are expected to play an important role in metallic systems. Following Ref.~\cite{Ma2025}, we refer to these contributions as metallic EO effects, which we discuss in greater detail in the following.


\subsubsection{A1. Metallic Magnetoelectric Electro-Optic Effects}
\label{subsubsec:A1}

A bias-induced correction to the gyrotropic magnetic coefficient, referred to as the magnetoelectric EO effect~\cite{magnetoelectric_electro-optical}, originates from the anomalous velocity correction by means of the prescription 
\begin{eqnarray}
\sigma_{B}^{\alpha\beta}(\omega) \rightarrow \sigma_{B}^{\alpha\beta}(\omega) + \epsilon_{\alpha\gamma\lambda}\chi_{E_0B}^{\lambda\beta}(\omega)E_0^{\gamma},
    \label{secII.eq3}
\end{eqnarray}
where~\cite{magnetoelectric_electro-optical}
\begin{eqnarray}
\chi_{E_0B}^{\alpha\beta}(\omega) = \displaystyle \frac{e^2}{\hbar} \frac{i\omega\tau}{i\omega\tau - 1}\sum_n \int \frac{d^3\textbf{k}}{(2\pi)^3}\left(-\frac{\partial f_{n\textbf{k}}^0}{\partial \epsilon_{n\textbf{k}}}\right)\Omega_{n\textbf{k}}^{\alpha}m_{n\textbf{k}}^{\beta}, \nonumber \\
    \label{secII.eq4}
\end{eqnarray}
It is possible to rewrite the full magnetoelectric coupling describing gyrotropic magnetic effects in a biased material in compact form by introducing the gyrotropic magnetic tensor $\textbf{K} = \hbar\sum_{n\textbf{k}}(-\partial f^0_{n\textbf{k}}/\partial \epsilon_{n\textbf{k}})\textbf{K}_{n\textbf{k}}$ and the magnetoelectric EO tensor $\textbf{G} = \sum_{n\textbf{k}}(-\partial f^0_{n\textbf{k}}/\partial \epsilon_{n\textbf{k}})\textbf{G}_{n\textbf{k}}$, with components related to $K^{\alpha\beta}_{n\textbf{k}} = v^{\alpha}_{n\textbf{k}}m^{\beta}_{n\textbf{k}}$ and $G^{\alpha\beta}_{n\textbf{k}} = \Omega^{\alpha}_{n\textbf{k}}m^{\beta}_{n\textbf{k}}$, respectively. The result is
\begin{eqnarray}
\boldsymbol{\sigma}_{B}(\omega)  = \displaystyle \frac{e}{\hbar}\frac{i\omega\tau}{i\omega\tau - 1}\left(\textbf{K} + e \textbf{F}_0 \cdot \textbf{G}\right),
    \label{Eq9}
\end{eqnarray}
where we have defined the a fully antisymmetric electric field tensor $\textbf{F}_0$, with components $F_0^{\alpha\beta} = -\epsilon_{\alpha\beta\gamma}E_0^{\gamma}$. We note that the contribution of the
$K$-tensor to the metallic optical conductivity is intimately connected to the so-called ``orbital Edelstein effect"~\cite{CoSi_Edelstein}, which refers to the generation of non-equilibrium orbital magnetization under applied electric fields. The static limit of this magnetoelectric response has been investigated in previous studies~\cite{pnas_Felser, chiral_optical, He2021}, which reported significant electrically induced orbital magnetization in topological chiral crystals belonging to SG198. In contrast, the contributions to the metallic optical conductivity arising from the $G$-tensor have not been reported since their initial theoretical prediction~\cite{magnetoelectric_electro-optical}. Accordingly, evaluating it for SG198 materials constitutes one of the central objectives of this work.

We now turn to the metallic EO effects arising from the Berry curvature dipole of Bloch electrons.


\subsubsection{A2. Metallic Electro-Optic Effects from the Berry Curvature Dipole }

In this section, we discuss relevant EO effects arising from the Berry curvature dipole of Bloch electrons~\cite{BCD_transistor, Ivo_LEO, magnetoelectric_electro-optical}. To linear order in the optical fields, the applied bias introduces two primary corrections to the optical conductivity, derived from Eqs.~(\ref{Eq4}) and (\ref{Eq5}). The corrections are captured through 
\begin{eqnarray}
\sigma_{E}^{\alpha\beta}(\omega) \rightarrow \sigma_{E}^{\alpha\beta}(\omega) + (\epsilon_{\alpha\gamma\lambda}\chi_{E_0E}^{\lambda\beta}(\omega) - \epsilon_{\alpha\beta\lambda}\tilde{\chi}_{E_0E}^{\lambda\gamma})E_0^{\gamma}, \nonumber \\
    \label{Eq10}
\end{eqnarray}
where 
\begin{subequations}
\begin{eqnarray}
\chi_{E_0E}^{\alpha\beta}(\omega) = \displaystyle \frac{\tilde{\chi}_{E_0E}^{\alpha\beta}}{1 - i\omega\tau}
    \label{secIII.eq7a}
\end{eqnarray}
\begin{eqnarray}
\tilde{\chi}_{E_0E}^{\alpha\beta} = \displaystyle \frac{e^3\tau}{\hbar} \sum_n \int \frac{d^3\textbf{k}}{(2\pi)^3}\left(-\frac{\partial f_{n\textbf{k}}^0}{\partial \epsilon_{n\textbf{k}}}\right)\Omega_{n\textbf{k}}^{\alpha}v_{n\textbf{k}}^{\beta}. 
    \label{secIII.eq7b}
\end{eqnarray}
\end{subequations}

We note that the momentum space integral in Eq.~(\ref{secIII.eq7b}) is simply the Berry curvature dipole up to a $\hbar$ factor. For instance, an integration by parts renders
\begin{eqnarray}
\displaystyle \hbar \int \frac{d^3\textbf{k}}{(2\pi)^3}\left(-\frac{\partial f_{n\textbf{k}}^0}{\partial \epsilon_{n\textbf{k}}}\right)\Omega_{n\textbf{k}}^{\alpha}v_{n\textbf{k}}^{\beta} = \int \frac{d^3\textbf{k}}{(2\pi)^3}f_{n\textbf{k}}^0 \frac{\partial \Omega_{n\textbf{k}}^{\alpha}}{\partial k_{\beta}}. \nonumber \\
    \label{Eq12}
\end{eqnarray}
Therefore, the optical conductivity can be fully expressed in compact form by introducing the Berry curvature dipole tensor  $\textbf{D} = \hbar\sum_{n\textbf{k}}(-\partial f^0_{n\textbf{k}}/\partial \epsilon_{n\textbf{k}})\textbf{D}_{n\textbf{k}}$, with components $D^{\alpha\beta}_{n\textbf{k}} = v^{\alpha}_{n\textbf{k}}\Omega^{\beta}_{n\textbf{k}}$ and the AC Drude tensor $\textbf{V} = \sum_{n\textbf{k}}(-\partial f^0_{n\textbf{k}}/\partial \epsilon_{n\textbf{k}})\textbf{V}_{n\textbf{k}}$, with components $V^{\alpha\beta}_{n\textbf{k}} = v^{\alpha}_{n\textbf{k}}v^{\beta}_{n\textbf{k}}$, as
\begin{eqnarray}
\boldsymbol{\sigma}_{E}(\omega)  = \displaystyle \frac{e^2\tau}{1 - i\omega\tau}\left(\textbf{V} - \frac{e}{\hbar ^2} \textbf{F}_0 \cdot \textbf{D}\right) + \frac{e^3\tau}{\hbar ^2}\textbf{D} \cdot \textbf{E}_0. \nonumber \\
    \label{Eq13}
\end{eqnarray}
We note that the contributions of the $D$-tensor to the metallic optical conductivity have been predicted in previous works~\cite{BCD_transistor, Ivo_LEO, magnetoelectric_electro-optical}. We also note that the zeroth order anomalous Hall conductivity contribution has not been included in the above equation due to our assumption of time-reversal symmetry. Together, Eqs.~(\ref{Eq1}), (\ref{Eq9}) and (\ref{Eq13}), account for the relevant optical responses arising from the coupling between optical fields, static bias and the wave function of Bloch electrons in inversion-broken, time-reversal symmetric systems. 

In the following, we apply the outlined approach to investigate the metallic EO effects in topological chiral crystals belonging to SG198. We start by detailing the tight-binding model that effectively captures the electronic behavior of these systems, particularly the presence of highfold semimetallic band crossings. These results will later be validated and complemented by direct first principles calculations in subsequent sections.

\section{III. Tight-binding model}
To investigate the metallic EO effects in nonmagnetic topological chiral crystals supporting multifold chiral fermions with large Chern numbers, we employ an eight-band tight-binding model respecting time-reversal and SG198 crystal symmetries~\cite{RhSi, SG198_SHE}. The non-trivial space group generators for these materials consist of $s_{2z} = \{ C_{2z} |\frac{1}{2} 0 \frac{1}{2} \} $, $s_{2y} = \{ C_{2y} |0 \frac{1}{2} \frac{1}{2}\} $, $ C_{3_{111}} = \{ C_{3_{111}} |0 0 0 \}$, and the cubic lattice vectors~\cite{pnas_Felser, RhSi, SG198_SHE}. Four s-orbitals are placed within the unit cell at atomic sites consistent with the 4a Wyckoff position and a spin-1/2 degree of freedom is considered at each site.

The next nearest-neighbor tight-binding model is~\cite{RhSi,SG198_SHE}
\begin{widetext}
\begin{equation}
\small
\begin{aligned} 
  &\mathcal{H}  =  t_s\tau _0 \otimes \beta _0 \otimes \sum_{i=1}^{3}  \cos \left( k_i \right) \sigma_0     \\
 &+ t_c ^{(1)} [ \cos \left(\frac{k_x}{2}\right) \cos \left(\frac{k_y}{2} \right) \tau _0 \otimes \beta _x \otimes \sigma _0  +  \cos \left(\frac{k_y}{2} \right) \cos \left( \frac{k_z}{2} \right) \tau _x \otimes \beta _x \otimes \sigma _0 + \cos \left(\frac{k_z}{2} \right) \cos \left(\frac{k_x}{2}\right) \tau _x \otimes \beta _0 \otimes \sigma _0] \\
 &+ t_c^ {(2)} [\cos\left(\frac{k_x}{2}\right) \sin\left(\frac{k_y}{2}\right) \tau _z \otimes \beta _y \otimes \sigma _0 +  \cos \left(\frac{k_y}{2}\right) \sin \left(\frac{k_z}{2}\right) \tau _x \otimes \beta _y \otimes \sigma _0 +  \cos \left(\frac{k_z}{2}\right) \sin\left(\frac{k_x}{2}\right) \tau _y \otimes \beta _0 \otimes \sigma _0]\\
 &+ \lambda _1 [\cos \left(\frac{k_x}{2}\right) \cos \left(\frac{k_y}{2} \right) \tau _0 \otimes \beta _y \otimes \sigma _z + \cos \left(\frac{k_y}{2}\right) \cos \left(\frac{k_z}{2}\right) \tau _y \otimes \beta _x \otimes \sigma _x + \cos\left(\frac{k_z}{2}\right) \cos \left(\frac{k_x}{2} \right) \tau _y \otimes \beta _z \otimes \sigma _y ]\\
 &+ \lambda _2 [\sin \left(\frac{k_x}{2}\right) \sin \left(\frac{k_y}{2} \right) \tau _z \otimes \beta _y \otimes \sigma _x + \sin \left(\frac{k_y}{2}\right) \sin \left(\frac{k_z}{2}\right) \tau _x \otimes \beta _y \otimes \sigma _y + \sin\left(\frac{k_z}{2}\right) \sin \left(\frac{k_x}{2} \right) \tau _y \otimes \beta _0 \otimes \sigma _z ]\\
 &+ \lambda _3 [\sin \left(\frac{k_x}{2}\right) \cos \left(\frac{k_y}{2} \right) \tau _0 \otimes \beta _x \otimes \sigma _x + \sin \left(\frac{k_y}{2}\right) \cos \left(\frac{k_z}{2}\right) \tau _x \otimes \beta _x \otimes \sigma _y + \sin\left(\frac{k_z}{2}\right) \cos \left(\frac{k_x}{2} \right) \tau _x \otimes \beta _0 \otimes \sigma _z ],
\end{aligned}
\normalsize
\end{equation}
\end{widetext}

where $t_s$, $t_c ^{(1)}$, and $t_c ^{(2)}$ are non-SOC hopping terms and the $\lambda _i$'s account for SOC between nearest neighbors. The $\tau$'s and $\beta$'s operate on the space of orbitals whereas the $\sigma$'s operate on the spin. For the model based calculations we set $t_s = 0.16eV$, $t_c ^{(1)} = 0.55eV$, $t_c ^{(2)} = -0.76eV$, $\lambda _1 = -0.03eV $, $\lambda _2 = 0.01eV $, and $\lambda _3 = -0.04$, which were previously determined from first principles based calculations of RhSi~\cite{RhSi}.

The energy spectrum of this model along a high symmetry path of the cubic Brillouin zone (BZ) is shown in Fig. 1a. Without SOC ($\lambda _i = 0$), but accounting for time-reversal symmetry and spin, the "little" co-group at $\Gamma$ has irreducible representations of dimension two and six, whereas at $R$ only eight-dimensional irreducible representations occur~\cite{Slager,Bradlyn2017, Bradlyn2,chiral_crystals, RhSi}. The nonrelativistic energy spectrum consists of a twofold (2f) node $2.2 eV$ above a sixfold (6f) one ($E = -0.07 eV$) at $\Gamma$ and a single eightfold (8f) node ($E = -0.48eV$) at $R$~\cite{SM}. SOC splits the 6f node at $\Gamma$ into fourfold (4f) and 2f nodes and the 8f one at $R$ is split into 6f and 2f~\cite{RhSi,SM}. The spectrum near $E = 0 eV$ at $\Gamma$ features a 4f chiral fermion with topological charge $C = -4$ about $0.1eV$ above a $C = -1$ Kramers-Weyl point (KWP) (Fig. 1b)~\cite{SM}. Near $E = -0.47 eV$ at $R$ there is a 6f chiral fermion with $C = +4$ (Fig. 1c)~\cite{SM}. Because the EO responses discussed in this work strongly depend on the nature of the momentum-space Berry curvature and orbital magnetic moment textures on the Fermi surface, we calculate these quantities for electron pockets formed by the uppermost bands near the $\Gamma$ and $R$ highfold crossings (Fig. 1d)~\cite{SM}. Both Berry curvature and orbital moment display a monopole-like texture, which can potentially lead to significant EO responses~\cite{magnetoelectric_electro-optical, pnas_Felser}. We explore this by calculating the energy dependence of \textbf{D}, \textbf{K}, and  \textbf{G} tensors in SG198 topological crystals.

\begin{figure}[h]
     \centering
          \includegraphics[width=\linewidth]{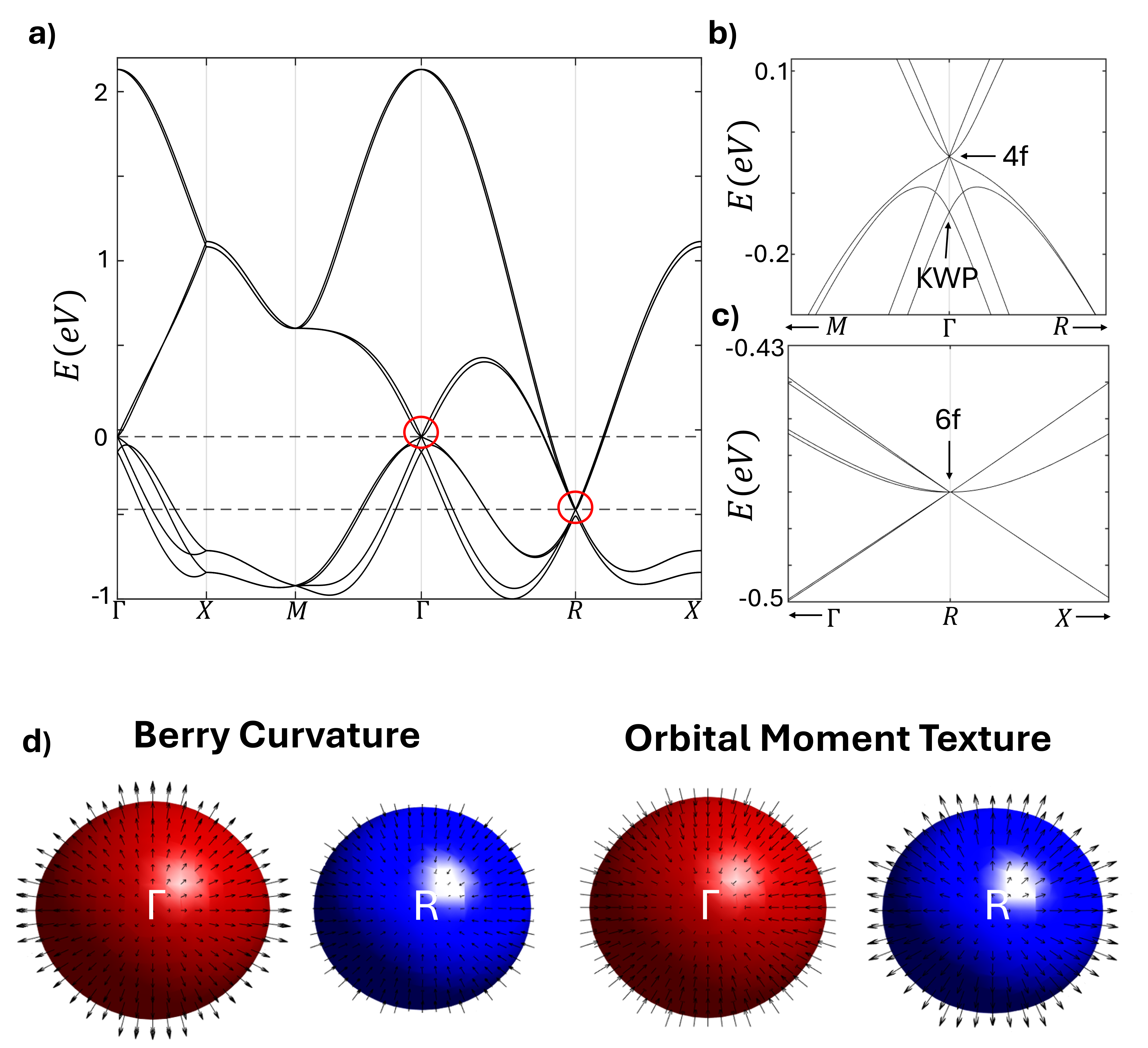}
      \caption{SG198 tight-binding model energy spectrum, Berry curvature, and orbital moment textures. a) Energy spectrum along high symmetry path of cubic Brillouin zone (BZ). Horizontal dashed lines indicate energies at which highfold degeneracies occur (red circles). b) Energy spectrum near fourfold (4f) fermion at $\Gamma$ with $C = -4$. $C = -1$ Kramers-Weyl point (KWP) appears $0.1eV$ below 4f fermion. c) Energy spectrum near sixfold (6f) fermion at $R$ with $C = +4$. d) Berry curvature and orbital moment textures of electron pockets near highfold nodes. Fermi surfaces correspond to topmost bands in b) and c) and are taken $30m$eV above each highfold node~\cite{SM}.}
\end{figure}


A detailed symmetry analysis of \textbf{D}, \textbf{K}, and \textbf{G} tensors in SG198 indicates a diagonal shape with a single independent coefficient for each tensor type (denoted $D$, $K$, and $G$, respectively)~\cite{SM}. The trace of \textbf{D} in SG198, which is related to the divergence of the momentum space Berry curvature, can be written as (see Eq. 12)
\begin{eqnarray}
Tr[\textbf{D}] = \sum_{n} \int \frac{d^3\textbf{k}}{(2\pi)^3}f^0 _{n \textbf{k}}\nabla_{\textbf{k}} \cdot \boldsymbol{\Omega}_{n \textbf{k}} 
= 3D(E)
\label{Eq15}
\end{eqnarray}
where $D$ is Fermi level, $E$, dependent. The absence of Berry curvature divergences leads to $Tr$[\textbf{D}] = 0, which would preclude the existence of nonvanishing \textbf{D} in SG198 crystals~\cite{Ivo_Te}. An improper operation (e.g. spatial inversion), $g$, maps a generic point in the BZ onto another $\textbf{k} \rightarrow g\textbf{k}$ and a point source (sink) of Berry curvature will be mapped onto a sink (source). If $g$ is a symmetry, these nodes are energetically aligned, which suggests a traceless \textbf{D} for all energies~\cite{Ivo_Te, Felser2}. However, chiral fermions of opposite topological charge pinned at the TRIMs are not connected by symmetry in chiral crystals, so there can be an energetic misalignment~\cite{RhSi, Pshenay-Severin_2018}. Thus, at fixed $E$ the k-space integration of the Berry curvature divergence weighted by the Fermi function may be nonzero, potentially permitting nonzero \textbf{D}~\cite{SM}. Previously assumed to vanish~\cite{Ivo_Te, Felser2}, below we predict the existence of \textbf{D} in SG198 crystals and highlight the contributions from the highfold fermions at $\Gamma$ and $R$ to $D$, $K$, and $G$.

\begin{figure}[b]
     \centering
          \includegraphics[width=\linewidth]{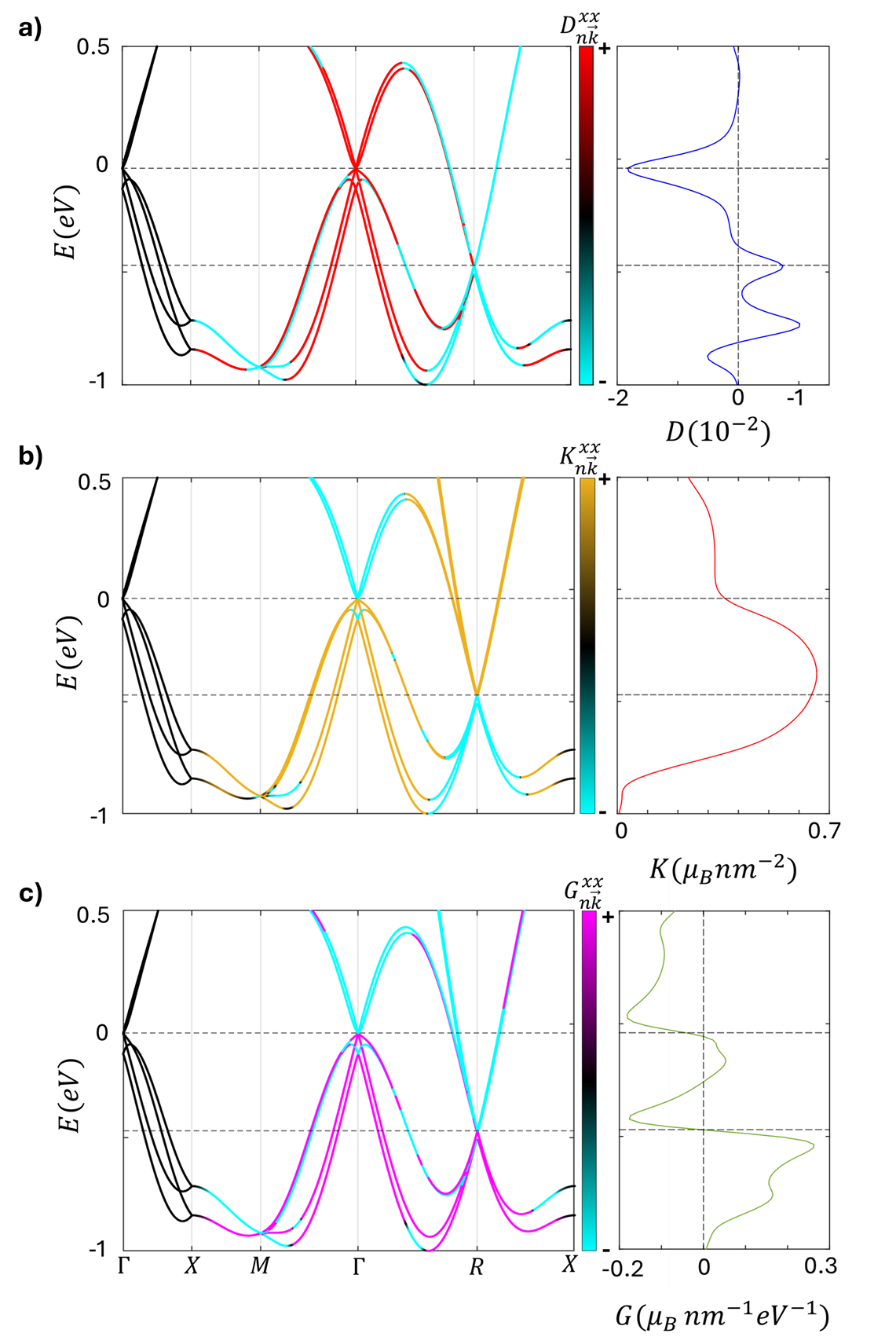}
      \caption{Tight-binding model Berry curvature dipole (\textbf{D}), gyrotropic magnetic (\textbf{K}), and magnetoelectric EO (\textbf{G}) coefficients. a) $D$-resolved energy spectrum and energy dependent $D$, b) $K$-resolved energy spectrum and energy dependent $K$, c) $G$-resolved energy spectrum and energy dependent $G$. For these calculations, the temperature and damping term are assumed to be 300K and 25meV, respectively. Dashed black horizontal lines indicate energetic positions of highfold fermions at $\Gamma$ and $R$. Note: The color scale for each tensor type is chosen to highlight the sign of the contributions and not their magnitudes. For magnitude comparisons, please see Supplementary Materials~\cite{SM}.}
\end{figure}


The energy dependence of $D$ for our tight-binding model is displayed in the right panel of Fig. 2a where we observe a valley and a peak near the energetic positions (dashed black lines) of the highfold fermions at $\Gamma$ and $R$, respectively. To elucidate the origin of these features, we calculate the $D$-resolved energy spectrum (Fig. 2a left panel), which shows that $D^{xx} _{n\textbf{k}} = v ^{x} _{n\textbf{k}}\Omega ^{x} _{n\textbf{k}}$ is nonzero primarily for bands emanating from the highfold topological band crossings at $\Gamma$ and $R$. We find that the largest contributions to $D$ come from states near these topological nodes~\cite{SM}. Effective $\textbf{k} \cdot\textbf{p}$ models, which accurately describe the bands that form the highfold crossings at $\Gamma$ and $R$, indicate that $D$ is constant above and below each of these nodes~\cite{SM, chiral_optical}. However, the presence of time-reversal symmetry requires that the Berry curvature be an odd function of the crystal momentum, $\boldsymbol{\Omega}_{n(-\textbf{k})} = -\boldsymbol{\Omega}_{n\textbf{k}}$. Consequently, the Berry curvature must vanish at the TRIMs in the vicinity of which the Berry curvature is large due to the presence of topologically charged point nodes~\cite{SM}. The result is an abrupt change in the Berry curvature near these points, which produces large $D^{xx}_{n\textbf{k}}$ (see Eq. 12)~\cite{SM} leading to the formation of the valley and peak observed in the right panel of Figure 2a. We note that the presence of an $s_{2y}$ symmetry in SG198 enforces the vanishing of $\Omega ^x _{n\textbf{k}}$ and $D^{xx}_{n\textbf{k}}$ along the $\Gamma$-$X$ high-symmetry line~\cite{SM}(see left panel of Figure 2a). 

Next, we calculate the energy dependence of $K$ and investigate its contributions by plotting the $K$-resolved energy spectrum (Fig. 2b). On account of its axial nature, $m^{x}_{n\textbf{k}}$ transforms identically to the Berry curvature under time-reversal and crystal symmetries. Thus, $m^{x}_{n\textbf{k}}$ also vanishes at the TRIMs and along certain high-symmetry lines~\cite{SM}. As was the case for $D^{xx}_{n\textbf{k}}$, $K^{xx}_{n\textbf{k}} = v ^{x} _{n\textbf{k}}m ^{x} _{n\textbf{k}}$ is nonzero for bands that form the highfold fermions at $\Gamma$ and $R$ and is largest for states near these band crossings~\cite{SM}. Our analytical results indicate that $K$ changes from a $ \beta(E-\epsilon _{\textrm{node}}) $ type energy dependence to  $ \alpha \sqrt{E-\epsilon _{\textrm{node}}} + \beta(E-\epsilon _{\textrm{node}})$ ($\alpha , \beta \in \mathbb{R}^+$)  across the 6f node at $R$, whereas across the 4f node at $\Gamma$, $K$ changes between $\propto \pm|E-\epsilon _{\textrm{node}}|$~\cite{SM}(right panel of Fig. 2b), which were previously predicted by Flicker, et al.~\cite{chiral_optical}.

Lastly, we calculate the energy dependent $G$ and plot the $G$-resolved energy spectrum (Fig. 2c). $G^{xx}_{n\textbf{k}}$ involves the product of $m^{x}_{n\textbf{k}}$ and $\Omega^x_{n\textbf{k}}$, so it must also be zero at the TRIMs and along specific high-symmetry lines~\cite{SM}. The left panel of Fig. 2c shows that the  bands emanating from the highfold topological nodes are responsible for the contributions to  $G$, the largest of which come from states near these fermions~\cite{SM}. Our analytical results of the energy dependence of $G$ around the highfold nodes indicate that $G$ changes between $ \propto \pm1/|E-\epsilon _{\textrm{node}}|$ type dependence across each node, which is observed in the right panel of Fig. 2c~\cite{SM}. To corroborate our tight-binding results, we now present first principles based calculations for 37 experimentally observed SG198 topological crystals.   

\section{IV. First Principles Calculations }

In order to connect our tight-binding model results with real bulk compounds, we studied the optical response tensors discussed above for 37 non-magnetic crystals from SG198 using density functional theory (DFT) based calculations~\cite{QE,SM}. For the purpose of calculating these responses, we projected the DFT Bloch wavefunctions onto a set of maximally localized Wannier functions to obtain a realistic tight-binding description of the bands around the Fermi level~\cite{W90,SM}. To illustrate a one-to-one correspondence between our tight-binding model and first principles based calculations, we designate RhSi as our material representative. Its crystal structure has eight atoms in the unit cell with Rh and Si forming opposite rotating chains along the [111] direction (see Fig. 3a).  The relativistic band structure at $\Gamma$ supports a 4f chiral fermion $0.08$eV above the Fermi level whereas a 6f chiral fermion at $R$ sits about $0.5$eV below it (Fig. 3)~\cite{RhSi}. The right panels of Fig. 3b-d present the energy dependent $D$, $K$, and $G$ and we next describe the sources of the displayed features.

\begin{figure}[H]
     \centering
          \includegraphics[width=\linewidth]{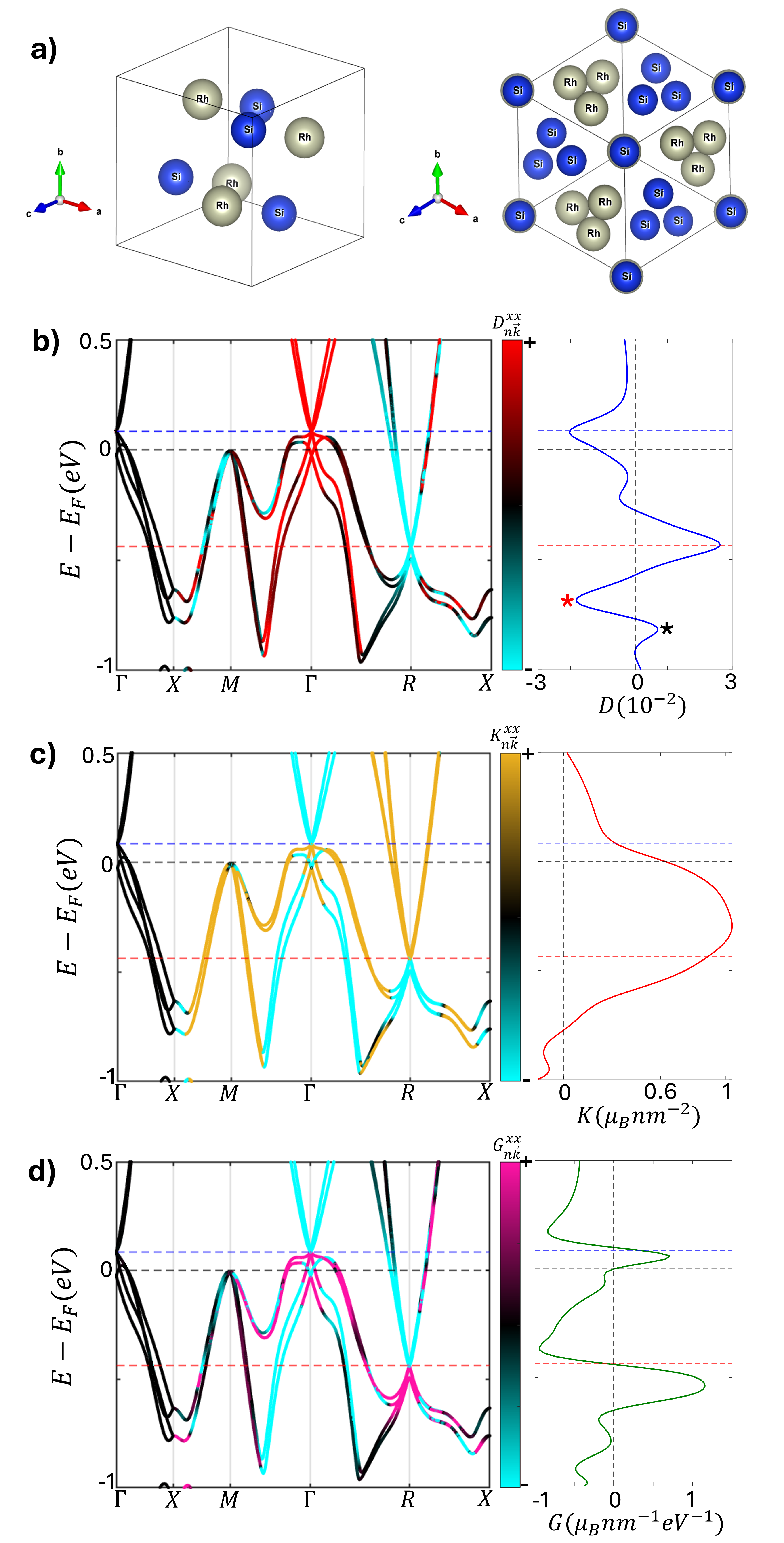}
      \caption{Structure of RhSi and energy dependent optical response tensors. a) Crystal structure of RhSi.  b) $D$-resolved band structure and associated energy dependent $D$. c) $K$-resolved band structure and associated energy dependent $K$. d) $G$-resolved band structure and  associated energy dependent $G$. For calculation details please see Supplementary Materials~\cite{SM}. Dashed blue and red horizontal lines indicate energetic positions of highfold fermions at $\Gamma$ and $R$, respectively. Note: The color scale for each tensor type is chosen to highlight the sign of the contributions and not their magnitudes. For magnitude comparisons, please see Supplementary Materials~\cite{SM}.}
\end{figure}
Well below the highfold nodes the bands mainly contribute positive $D^{xx}_{n\textbf{k}}$~\cite{SM} (left panel of Fig. 3b) and a small positive peak (black asterisk) is observed. Above this peak a valley appears (red asterisk), which is a consequence of relatively large negative contributions coming from states along the $\Gamma$-$R$-$X$ high symmetry line~\cite{SM}. The large peak and valley centered near the energetic positions of the highfold crossings (dashed blue and red horizontal lines) are the result of the sudden vanishing of the Berry curvature at the TRIMs as described in the previous section. 

 At low energies large negative $K^{xx}_{n\textbf{k}}$ from bands along the $M$-$\Gamma$ line (left panel Fig. 3c) originating from the KWP at $\Gamma$ give rise to negative $K$(right panel Fig. 3c)~\cite{SM}. As the energy is increased, additional positive contributions from bands along the $X$-$M$ and $R$-$X$ high symmetry lines lead to a sign change in $K$. A large peak in $K$ is observed around $0.285$eV below the Fermi level largely on account of intense positive $K^{xx}_{n\textbf{k}}$ arising from electron pockets centered above the 6f crossing at $R$~\cite{SM}. Near the Fermi level, a decrease in the number of bands with positive $K^{xx}_{n\textbf{k}}$ and negative contributions from states above the KWP at $\Gamma$ lead to the displayed rapid decrease in $K$. At higher energies, large negative contributions from states near the 4f node at $\Gamma$ cause a further decrease in $K$ and an eventual sign change~\cite{SM}. 
About 0.54eV below the Fermi level there is a peak in $G$ (right panel of Fig. 3d), which is a consequence of the strong positive $G^{xx}_{n\textbf{k}}$ coming from states immediately below the 6f node at $R$~\cite{SM}. A large valley is observed above this peak, which arises from negative $G^{xx}_{n\textbf{k}}$ from electron pockets centered above the 6f crossing at $R$~\cite{SM}. Above the Fermi level, a second positive peak appears which stems from positive $G^{xx}_{n\textbf{k}}$ associated with hole pocket states below the 4f node at $\Gamma$~\cite{SM}. As the energy is increased across the 4f node, negative contributions from the electron pockets at $\Gamma$ bring about a sign change. We note that the largest contributions to $D$, $K$, and $G$ come from states near the $C = \pm4$ topological nodes~\cite{SM} and the overall behaviors of the energy dependent $D$, $K$, and $G$ of Fig. 3 are consistent with the results of the previous section, further validating our tightbinding model and $\textbf{k} \cdot\textbf{p}$ analysis. In what follows we make comparisons between these responses for electrically biased SG198 materials. 

The conductivities of Eqs. (9) and (13) suggest that $D$- and $G$-dependent metallic EO effects are dominant above specific bias fields. We define the critical fields as $E^E_{0_c} =\frac{\hbar ^2}{e}|\frac{V}{D}|$ and $E^B_{0_c} = \frac{1}{e}|\frac{K}{G}|$. Fig. 4 displays Fermi level critical fields for 37 topological crystals from SG198. Our material representative, RhSi, has critical fields $E^E_{0_c} \approx 5 \times 10^{9} \frac{V}{m} $ and $E^B_{0_c} \approx 3 \times 10^{11} \frac{V}{m}$, which indicates that $D$-dependent EO effects are more accessible than those related to $G$. BeAu has the lowest possible $E^B_{0_c}$ of this set of materials at $8 \times 10^5 V/m$ as a consequence of the relatively low $|K|$ near the Fermi level~\cite{SM}. More importantly, $E^E_{0_c} \approx 18 \times 10^{9} \frac{V}{m}$ is well above  $E^B_{0_c}$ for this material. Thus, as a function of electric bias, the onset of $G$-dependent magnetoelectric EO effects~\cite{magnetoelectric_electro-optical} in this material precede those associated with the BCD, making BeAu the most experimentally viable candidate for probing these novel EO properties. 


\begin{figure}[H]
     \centering
          \includegraphics[width=\linewidth]{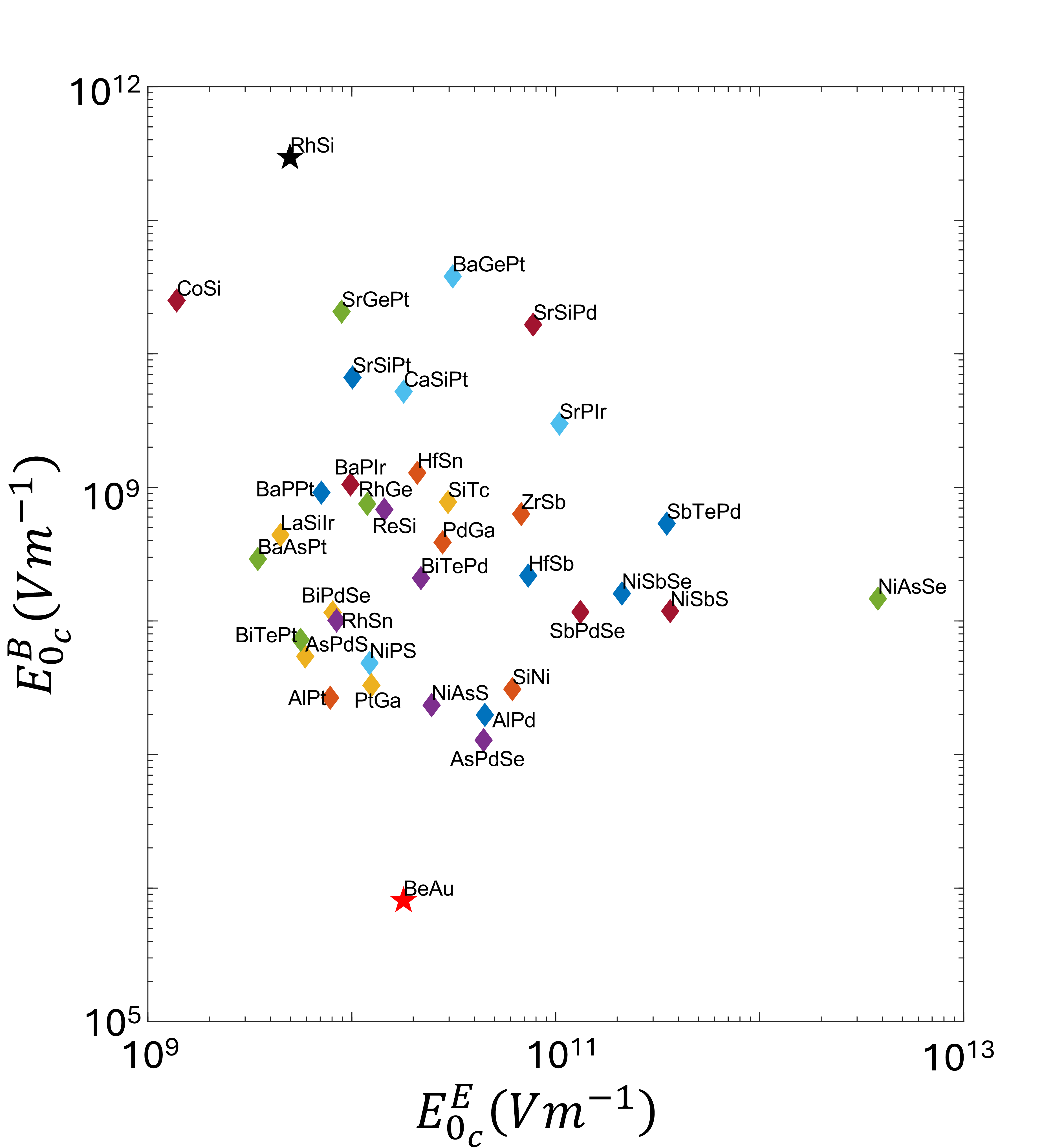}
      \caption{Critical fields for EO response of SG198 materials. $E^B_{0_c} = \frac{1}{e}|\frac{K}{G}|$ measures the onset of the novel EO effects associated with $G$ whereas $E^E_{0_c} =\frac{\hbar ^2}{e}|\frac{V}{D}|$ does so for EO effects related to $D$. Fermi level $D$,  $K$, $G$, and $V$ are used to calculate these critical fields~\cite{SM}.}
\end{figure}


\section{V. Conclusions}
Using a tight-binding model and density functional theory (DFT) based calculations, we explored the metallic electro-optic (EO) effects in topological chiral crystals belonging to space group 198 (SG198) with an emphasis on the contributions arising from highfold chiral fermions located at the $\Gamma$ and $R$ points of the Brillouin zone. Contrary to previous assumptions, our findings indicate that the Berry curvature dipole is nonzero in these systems, arising as a result of the energy offset between topologically charged nodes of opposite chirality. Among the 37 SG198 materials examined, BeAu stands out as a strong candidate for realizing the predicted magnetoelectric EO effects under an experimentally accessible electric field of $8 \times 10^5$ V/m.

\textit{Acknowledgments}. C.O.A., D. S. and T. L. acknowledge support from Office of Naval Research MURI grant N00014-23-1-2567.

\bibliography{manuscript.bib}
\end{document}